# An Entire Renal Anatomy Extraction Network for Advanced CAD During Partial Nephrectomy

Nan Ma, Ying Yang, Dongkai Zhou

**Abstract:** Partial nephrectomy (PN) is common surgery in urology. Digitization of renal anatomies brings much help to many computer-aided diagnosis (CAD) techniques during PN. However, the manual delineation of kidney vascular system and tumor on each slice is time consuming, error-prone, and inconsistent. Therefore, we proposed an entire renal anatomies extraction method from Computed Tomographic Angiographic (CTA) images fully based on deep learning. We adopted a coarse-to-fine workflow to extract target tissues: first, we roughly located the kidney region, and then cropped the kidney region for more detail extraction. The network we used in our workflow is based on 3D U-Net. To dealing with the imbalance of class contributions to loss, we combined the dice loss with focal loss, and added an extra weight to prevent excessive attention. We also improved the manual annotations of vessels by merging semi-trained model's prediction and original annotations under supervision. We performed several experiments to find the best-fitting combination of variables for training. We trained and evaluated the models on our 60 cases dataset with 3 different sources. The average dice score coefficient (DSC) of kidney, tumor, cyst, artery, and vein, were 90.9%, 90.0%, 89.2%, 80.1% and 82.2% respectively. Our modulate weight and hybrid strategy of loss function increased the average DSC of all tissues about 8-20%. Our optimization of vessel annotation improved the average DSC about 1-5%. We proved the efficiency of our network on renal anatomies segmentation. The high accuracy and fully automation make it possible to quickly digitize the personal renal anatomies, which greatly increases the feasibility and practicability of CAD application on urology surgery.

## I. Introduction

Comparing to radical nephrectomy, partial nephrectomy (PN) has no statistically significant increases in intraoperative complications and postoperative adverse events, but brings more preservation of renal function which greatly benefit to patients[1]. The precise excision range and accurate arterial clamping, will improve the quality of PN with less warm ischemia time (WIT), reducing blood loss (EBL) and shorter operative time. However, the variety of spatial relationship between renal tumor and its artery supply proposes challenges to urologist surgeon. They must design the operative plan for every patient.

With the rapid development of computer technology, tons of computer-aided diagnosis (CAD) techniques are emerged gradually. The superiority of CAD applications during partial nephrectomy have been proven in most researches. By comparison, Wang et al. found that the preoperative three-dimensional (3D) reconstruction technique ensured more preserved renal parenchymal mass and shorter warm ischemia time when facing complex renal tumors during laparoscopic partial nephrectomy (LPN) [2]. Su et al. applied the Augmented Reality (AR) based image-guided surgery system on LPN. The projection of 3D-registration model on the real-time stereoscopic video makes it possible to allow urologist surgeon to see beneath the veil of kidney surface [3]. But its accuracy remains uncertain due to the immature registration algorithm and human error of manual segmentation. After studying other more researches in relevant areas, we found a consensus that the performances of all CAD techniques mentioned in their respective researches, rely on the accuracy of renal anatomies digitization. However, the manual delineation of kidney vascular system and tumor on each slice is time consuming, error-prone, and inconsistent. Therefore, we turn to advanced computer science for help.

A large amount of automatic or semi-automatic organ segmentation extraction algorithm were presented in the last two decades. Lin et al. combined the region-based and model-based methods, successfully developing an automatic kidney segmentation system from abdominal computed tomography (CT) images [4]. Yang et al. presented a coarse-to-fine strategy in segmentation of kidney by using multi-atlas image registration to achieve great accuracy in detail [5]. Song et al. proposed kernel fuzzy C-means algorithm with spatial information (SKFCM) algorithm to refine Grow Cut (IGC) algorithm kidney segmentation output [6].

Deep Learning technique is developing remarkably in recent years, and it already showed significant achievement in many organ segmentations challenges. Akbari et al. introduced Wavelet-based support vector machines (W-SVMs) application on renal extraction by analysis Magnetic Resonance image (MRI) texture feather, and developed a weight functions to balance the Wavelet-based, intensity-based, and model-based label for better result [7]. Sharma et al. presented automated segmentation of Autosomal Dominant Polycystic Kidney Disease (ADPKD) kidneys using fully convolutional neural networks(CNNs) to facilitates fast measurements of kidney volumes [8]. Gibson et al. developed a registration-free segmentation algorithm for eight abdominal organs simultaneously based on deep-learning [9].

Due to the volume image noise and delicate vessel shapes, vascular system segmentation from CT and MRI is much harder than organ. Still, many advanced algorithm studies have implemented satisfactory results. Bauer et al. constructed vessel shape prior for graph cuts segmentation of 3D vessels by applying a image filter and height ridge traversal method [10]. Wang et al. used a Tensor-based Graph-cut method improving accuracy in tiny blood vessels segmentation. The method has advantage in renal vessel extraction task, comparing to other vessel segmentation algorithms [11]. Deep learning also showed remarkable result in this aera. Huang et al. introduced a robust liver vessel extraction algorithm implanted end-to-end image segmentation with few training samples via a dense convolutional network together with post-processing [12]. Huang et al. also proposed Deep Learning Neural Networks application in Coronary Artery Segmentation from Computed Tomographic Angiographic (CTA) Images, and realized great result [13].

Fully convolutional networks (FCNs) are mentioned in last two studies above. It is a widely used network in 2D/3D medical segmentation practices, providing precise pixel-wise/voxel-wise prediction [14]. Recently, many improvements to FCNs architectures have been proposed. U-Net is one of the most unignorable extended version of FCNs [15], characterized by a series of symmetric convolutional layers and connection between the down-sampling path and the up-sampling path, forming U-shaped structure.

Credit to its excellent performance, U-Net stood out from other FCNs. Gradually it become standard network for any biological segmentation problems. More state-of-the-art U-Net-like architectures emerged. Alom et al. proposed Recurrent Residual Convolutional Neural Network (R2U-Net). It has the both advantages of recurrent unit and residual unit, and showed better performance on many benchmark datasets[16]. Zhou et al. present UNet++ with a series of nested, dense skip pathways. The UNet++ is proved superior on both 2D and 3D medical segmentations. [17]

Despite the extensive usage, U-Net is restricted by hardware when handling volumetric data (3D data) like CT or MRI. Normally, 3D data is hundred times the size of 2D data. Graphic Processing Units (GPU) provide high-speed matrix operation for neural network computation. However, only high-end GPU cards offer large-capacity video memory. This led to hardware restriction of 3D U-Net application on GPU.

In this paper, we provide a fully deep learning based approach for entire renal anatomies segmentation from CTA images. We chose coarse-to-fine workflow to tradeoff between the volumetric sampling resolution and GPU memory consumption, by applying the same 3D U-Net architectures at two image sampling levels: low resolution and high resolution. We also set up many experiments of comparing different variables relevant to training to seek the best-fitting. At last, we proved that our algorithm can efficiently establish the relative spatially relationship between renal tumor and vessels in real physical space, which is useful to CAD process during precise PN surgeries.

## II. Methods

Our segmentation method consists of 2 stages with different sampling levels. Stage I, also deemed as "coarse stage", we roughly extract kidney from low resolution CTA (resampled from full resolution) by applying the 3D U-Net, aiming to locate the kidney and generate the region of interest (ROI). Then, we crop the full resolution CTA based on the ROI of kidney, which fully contains the target kidney and its tumor. Stage II, also deemed as "fine stage", we accurately classify the all kidney tissues and vascular system by the same architecture 3D U-Net used in stage I from high resolution CTA separately (stage IIA and stage IIB). **Fig. 1** shows the main workflow of our method.

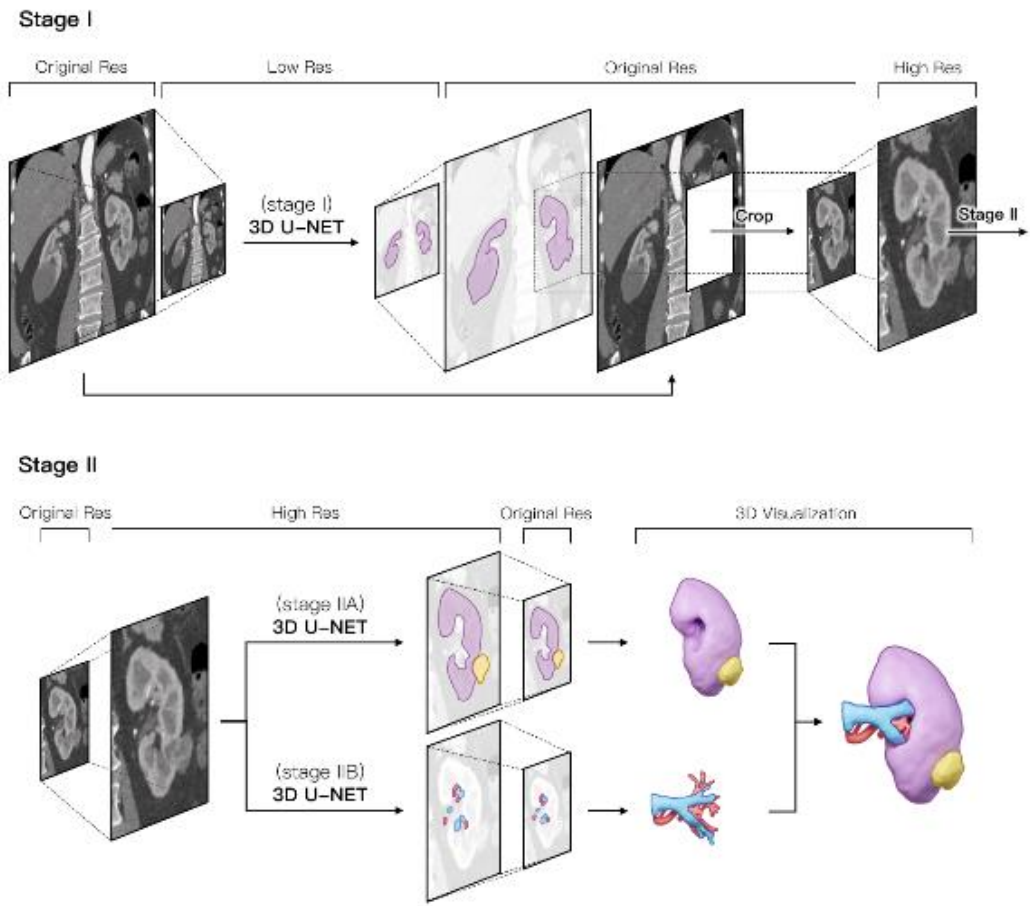

**Fig. 1** Stage I: down-sample the input image for stage I segmentation, then resize the prediction image to its original resolution for cropping. Stage II: up-sample the cropped full resolution CTA from stage I, then feed it to stage IIA and stage IIB model for kidney tissues and vascular system extraction separately. Finally, merge all predicted tissue together and send to 3D engine for visualization. The slice image represents volumetric data, rather than single slice.

## A. Training datasets

Different datasets are used in two stages. Considering the robustness of the network, we choose KiTS19 dataset (https://kits19.grand-challenge.org/) for stage I training. This available public dataset including 210 contrast-enhanced abdominal CT scans of kidney cancer patients with manual segmentation labels of kidney and tumor. The diversity of image resolutions, intensity distributions and anatomical structure ensures the accuracy of localization, which is the fundament of stage II prediction. The stage II dataset mainly from 3 sources: 20 cases were from KiTS19 dataset. The contrast-enhanced CT scans in KiTS19 dataset is various in phases. Therefore, we only selected those in arterial phase, also guaranteed the diversity of volume spacing and anatomical structure. Another 20 cases were from The Second Affiliated Hospital of Xi'an Jiao tong University. all patients in this cohort underwent PN for one or more kidney tumors at the Urological Surgery department of The Second Affiliated Hospital of Xi'an Jiao tong University between 2015 and 2019. Their preoperative CTA scans were collected. The last 20 cases were from The Second Affiliated Hospital of Zhejiang University School of Medicine. The inclusion criteria are same as source 2. Those 3 sources datasets were mixed and shuffled, then combined into one dataset for stage II training and testing.

Although KiTS19 dataset provided great number of manual segmented labels image data, but some of the cases have annoying artificial defects and annotation errors on medicine. Moreover, we want to distinguish the parenchyma, hilum, and vessel of the renal, but all those were considered as same label in KiTS19 annotation criterion. So, we decided to discard the label images that KiTS19 dataset provided, and only regarded as references.

The training segmentation labels of the 60 cases in stage II are manually annotated by two experienced clinic radiologists independently. We annotated 5 tissue classes into 2 label images. The First label image contains renal parenchyma (including cortex and medulla, also named as "kidney" below), tumors, and cysts, while the second contains renal arteries and veins. All annotation procedures were operated on 3D slicer (version 4.10.2), an open source software platform for medical image informatics, image processing, as well as three-dimensional visualization. Each tissue label is processed in 4 steps, we take the kidney class's annotation for example:

Firstly, the annotator drew one or more contours that containing all renal tissues excluding the collecting system and vessel in renal hilum on evenly spaced slices. Secondly, the slices in-between is generated by the "fill between slices" method that provided by 3D slicer to complete the work on every slices. Thirdly, the annotator checked the computed result and fixed the error region on every axis slice by slice. Finally, we apply gaussian blur method in 3D slicer to remove the obvious artifacts and jagged boundaries. For facilitating the tissue localization accuracy and exclusion of cysts from tumors, the annotators also used each case's attending radiologist's conclusions, surgical record, and pathology results as references.

In stage I dataset (210 samples), the pixel spacing varied from 0.43 to 1.04 mm, the slice thickness varied from 0.5 to 5.0 mm, and the slice number varied from 29 to 1059. In stage II dataset (60 samples), the pixel spacing varied from 0.74 to 0.96 mm, the slice thickness varied from 0.5 to 1.0 mm, and the slice number varied from 261 to 738.

### B. Optimizing vessel annotation

Reconstructing branching conduit structures like arteries or bronchus from volumetric images by human brain is not an easy job. Additionally, some low contrast aeras are awfully hard to be identified, adding more difficulties to it. Furthermore, the low-quality image such as serious noise and blurred boundaries will also influence image recognition. Thus, we cannot ask for more accuracy relied on manpower entirely, which has been proved by our practices.

As showed in **Fig. 2**, we found that there were many unlabeled vessels in stage IIB label images by comparing our computed result and original annotated dataset. It has been confirmed by experts that some unlabeled vessels extracted via algorithm are correct. We cannot just ignore these valuable annotations provided by artificial intelligence. An extra data preprocess step therefore was added before feeding to training model: we extracted all unlabeled vessels by subtracting manual annotation from calculated annotation for each class. Then, we selected those extra vessels based on medical information under the supervision of experts and merged them onto original label image.

However, this was also a great challenge for human in our practices. Because, our goal is extracting both arteries and veins, while they are unable to distinguish only based on voxel intensity when they are going further due to the low contract. Although, we can rely on the going trend of vessels, it still cannot be sure whether they are arteries or veins because of the thin structure. Under the circumstances, we stipulated that arties has a higher priority than veins. Those controversial tiny vessels were assumed as arteries since arteries accompanied closely with veins in renal columns, and we require more accuracy when segment artery.

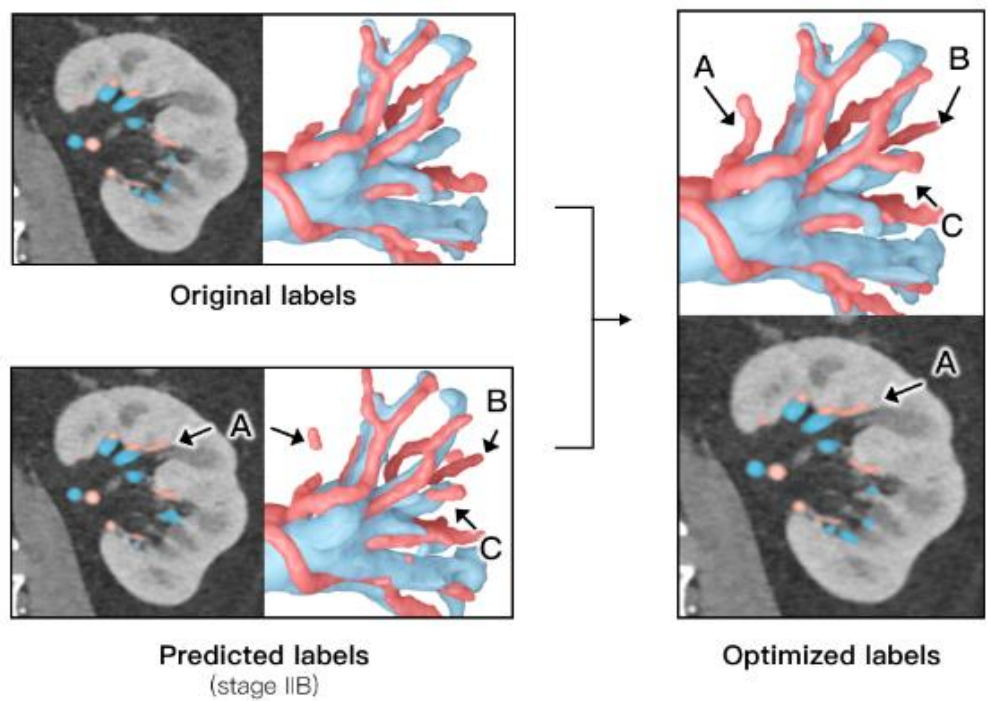

Fig. 2 A, B, C are the vessels that were not labeled in original image. We used the semi-trained models' prediction image to optimize the original annotations under supervision.

### C. Preprocessing

Before training, a series of image data preprocessing is carried out in both stages with different parameters.

As mentioned above, the 3D U-Net is great GPU memory consuming. reducing the size of the input image benefit the training process. Therefore, all data has been cropped to the bounding box of human tissue. We assume that Hounsfield Unit (HU) value below -200 is non-human region. Removing those area will not affect the result, but dramatically reduced computation. Additionally, the 60 cases in stage II are cropped to the kidney area based on the mask of annotations. After that, we got 120 images with their original resolution for training.

After cropping, we oriented the volumetric input data to Right-Anterior-Superior (RAS)

coordinate system, and uniformed the voxel spacing. Because of the multiple sources, there is no benchmark to ensuing the same orientation of images. In addition, the voxel spacing of data is also inconsistent, due to the difference of the CT scanner settings in clinical procedures. If we do not put those meta info (coordinate system and voxel spacing) into the network, it would not understand the spatial semantics in real physical space natively. Instead of transferring those meta info into the bottom of the U-Net architecture, which may increase the computation while training, we rather to fix it in data processing phase.

On the one hand, higher resolution resampling expands the image size, increasing the calculation burden. On the other hand, lower resolution resampling does cause potential data loss, as well as discontinuities effects. If the target tissue is about the size of spacing width, it might be wiped out through resampling. So, we decided to adopt the coarse-to-fine workflow: two different levels of resampling are used in two stages to solve the contradiction. In stage I, we want to cover the entire medical image as much as possible, paying less attention to the details of tissue except the locations. So, we set the uniformed voxel spacing to $2.4 \times 2.4 \times 3.0$ mm, resulting in a median resampled image shape of $166 \times 142 \times 136$ voxels. In stage II, we focus more on the fine detail of the tissues' edge, without missing any tiny structures. So, we set the uniformed voxel spacing to $0.7 \times 0.7 \times 1.0$ mm, resulting in a median resampled image shape of $157 \times 156 \times 142$ voxels.

Normalization is also involved in preprocessing. All CTA images were clipped to the [0.5, 99.5] percentiles, and applied z-score normalization based on the mean and standard deviation of their entire foreground voxel intensity values. Concretely, we clipped stage I cases to range [−79, 303], then subtract 100.2 and divide by 76.6. We clipped stage II cases to range [−69, 426], then subtract 137.5 and divide by 88.9. The nature of weight initialization to organ-specific value range facilitates the network training process.

### D. Network architecture

Our proposed network architecture is mainly based on 3D U-Net architectures, containing a down-sampling path for feature extraction and an up-sampling path for image-to-image segmentation. **Fig. 3** illustrates the overview of the architectures.

3D U-Net is one of the most used networks for volumetric medical image extraction tasks. In the vanilla 3D U-Net, each encoder block along down-sampling path consists of two padded convolutions. We replaced them with the increasing number of residual blocks as path going down to avoid degradation problem caused by the deeper network. Every residual block contains two convolutional layers (kernel size $3 \times 3 \times 3$), each followed by an instance normalization (IN) layer and a leaky rectified linear unit (Leaky ReLU) as an activation function. The shortcut connection of the residual is carried out before the last Leaky ReLU (see **Fig. 3**). After each down-sampling, the stack size of stacked residual blocks increased (from 1 to 5), as well as the feature channels (from 30 to 480). The decoder block along up-sampling path has only one residual block with the same structure as that in the encoder block. Instead of using max-pooling layer for down-sampling operations in standard U-Net, we also used residual block with stride 2 as the replacement. Thus, the down-sampling operations will not damage the tiny anatomies such as arterial branches, but also taking well care of the large anatomies. Those learnable parameters provided by the 2 strides residual block makes the down-sampling path much wiser than max-pooling function. We also designed a 2 × 2 × 2 up-convolution with stride 2 for up-sampling operation, followed by a skip connection from the encoder block with the same resolution before transferred into the decoders. The skip connection was built by the concatenation of the encoder output and the up-convolution output along feature channel. The last layer is a 3 x 3 x 3 convolutional layer with a specific number of feather channels according to the kinds of tissue we want to extract. Finally, a sigmoid or a softmax function is performed to generate the probability maps for tissue.

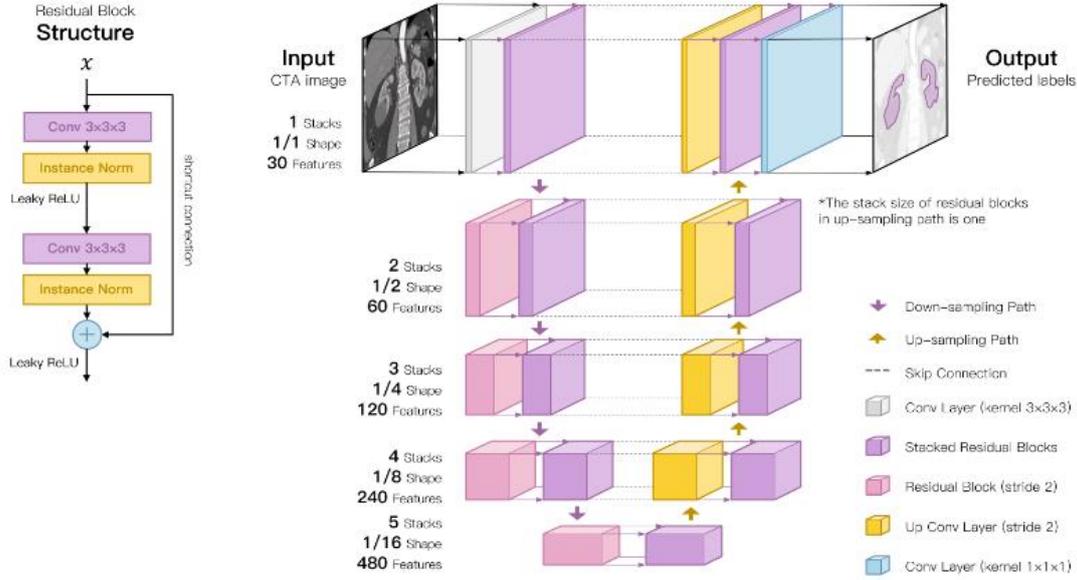

**Fig. 3** Our proposed network architecture. Before final output, we applied softmax or sigmoid function according to the number of classes. The stack size is all one in up-sampling path. The upper left is the structure of residual block used in stacked residual blocks.

## E. Loss function

We adopted the combination of dice loss and focal loss as our entire loss function when training the models.

$$L_{total} = L_{dice} + L_{focal}$$

The dice loss is derived from dice score coefficient (DSC), which is defined as the extent of the overlap between predicted probability and ground truth. It is calculated as follows:

$$L_{dice} = \sum_{c=0}^{C-1}(1 - DSC_c)$$

$$DSC_c = \frac{\sum_{n=1}^{N} p_{n,c} g_{n,c} + \epsilon}{\sum_{n=1}^{N} p_{n,c} + \sum_{n=1}^{N} g_{n,c} + \epsilon}$$

Where $DSC_c$ represents dice score coefficient of class $c$, $p_{n,c}$, $g_{n,c}$ represent predicted probability and ground truth of voxel $n$ for class $c$ respectively, $C$ is the total number of classes, $N$ is the total number of voxels, $\epsilon$ provides numerical stability of loss to prevent division by zero and was set to 1e-7.

DSC is one of the most widely accepted overlap criterions for medical segmentations. Under some circumstances, we need improve the sensitivity of prediction to expose more potential regions. But DSC lacks such modulating controls. We therefore employed an alternative of DSC named "Tversky similarity index" with two parameters $\alpha$ and $\beta$ to modify the weights of false negatives (FN) and the false positives (FP). The formula of DSC was transformed to this:

$$DSC_c = \frac{\sum_{n=1}^{N} p_{n,c} g_{n,c} + \epsilon}{\sum_{n=1}^{N} p_{n,c} g_{n,c} + \alpha \sum_{n=1}^{N}(1 - p_{n,c}) g_{n,c} + \beta \sum_{n=1}^{N} p_{n,c}(1 - g_{n,c}) + \epsilon}$$

Tversky similarity index is more like a generalization of DSC. They are mathematically equivalent when both $\alpha, \beta$ are 0.5. But with the parameters, we can slide between the FN and FP, which affects the emphasis of training direction. In experiments section, we will demonstrate how $\alpha$ and $\beta$ influence the prediction results.

The focal loss is also involved in the total loss. It is implemented as follows:

$$FL_c = -\frac{1}{N} \sum_{n=0}^{N} (1 - p_{n,c})^r g_{n,c} \log(p_{n,c})$$

$$L_{focal} = \sum_{c=0}^{C-1} FL_c$$

Where $\gamma$ is the focusing parameter to modulate the focus. We set the $\gamma$ to 2.

The focal loss can be regarded as an enhanced version of the cross-entropy (CE) loss. Tiny tissue extraction from background is always tougher than larger ones. As it showed in **Fig. 4**, the voxel area of cysts and vessels are much lower than kidney. The equation of CE determined that objects with smaller areas contribute less to the entire loss, which can easily result in classification deviation. The focal loss was designed to address this problem that came from the extreme imbalance contributions. It adopted a modulating factor $(1 - p_{n,c})^r$ to the original cross-entropy, promising extra attention on terribly classified regions resulted from less contribution[18].

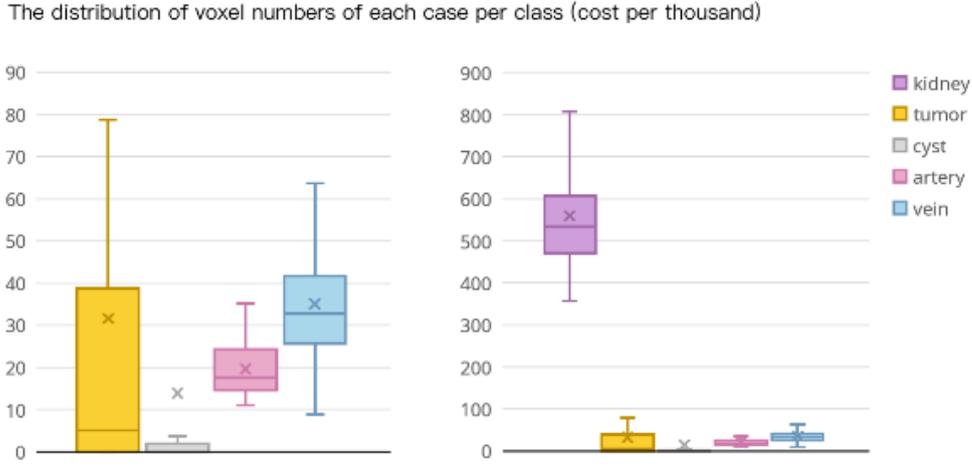

Fig. 4 The distribution of voxel numbers of each case per class. The background class is not presented because it takes up 90.7% of whole image averagely, significantly greater than other five classes. The left box plot excludes the kidney class for better display of small objects.

To resolve this imbalance issue even further, more effort was token. We propose to add an adjustable weight $w_c$ to both dice loss and focal loss for rebalancing. $w_c$ is inversely proportionate to the voxel-wise frequency of classes. With such assurance, less voxels class will receive more weight than others.

So far, the complete formula of the loss can be written as follows:

$$L_{total} = \sum_{c=0}^{C-1} w_c((1 - DSC_c) + FL_c)$$

The anther annoying issue is missing annotation label. It certainly hurts the stability of model training process. **Fig. 5** lists all classes' case-wise frequency: tumor and cyst are not always presented in cropped CTA image, as many patients' contralateral kidneys are health as normal. To handle this problem, we wiped out the missing annotations' weight and redistribute the rest in certain training batch. But this move also increased the rest classes' contribution to loss which we do not want. So, another factor was added to $w_c$ for correcting. This factor is inversely proportionate to the case-wise frequency of classes. Synthesis above, the weight $w_c$ is calculated as follows:

$$m_c = \begin{cases} 1, & if \ G_c > 0 \\ 0, & otherwise \end{cases}$$

$$w_c = \frac{m_c fc_c fv_c}{\sum_{i=0}^{C-1} m_i fc_i fv_i}$$

Where $m_c$ represents the mask of the weights. If $G_c$, the maximum of ground truth in class $c$, is on larger than 0 which means class $c$ annotation is missing, $m_c$ will be 0 to remove the contribution. $fc_c$, $fv_c$ represent case-wise frequency and voxel-wise frequency of class $c$ respectively.

With this super enhanced weight $w_c$, all class will be treated fairly during training. Any speculative behavior caused by excessive attention will be prevented systematically.

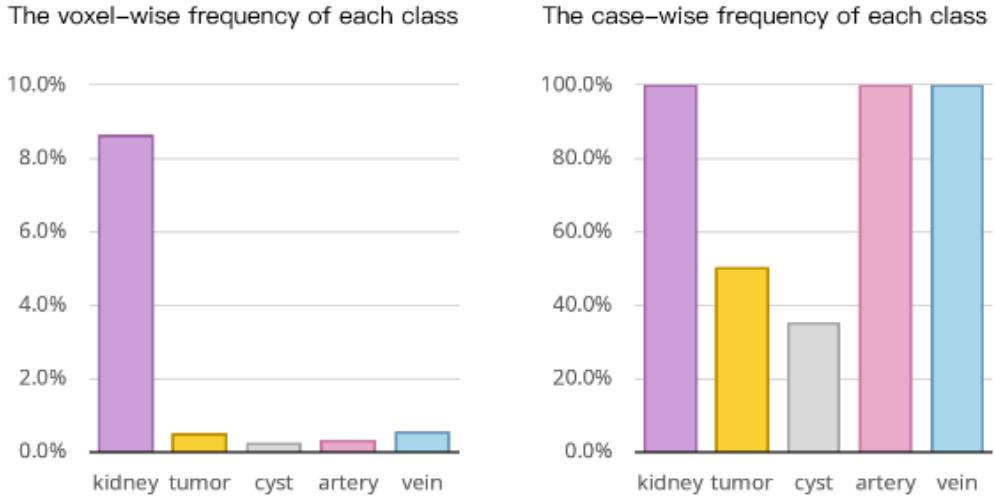

**Fig. 5** The voxel-wise frequency and the case-wise frequency per class.

## F. Training platform and evaluation metrics

Our proposed algorithm was implemented using Python 3.6.1 and Pytorch framework 1.4.1. Training was done on a computer with an AMD Ryzen™ 9 3900X (3.8GHz 32.0 GB RAM) and an NVIDIA Geforce™ GTX 1080Ti-11G (11 GB VRAM). In all experiments, we trained the models for 1000 epochs at most. Each epoch is defined as the iteration over 200 batches. The average training time was about 72 hours. We set the different patch size based on the median shape of their respective training dataset in different stage, to ensure retaining enough necessary information for learning while considering the maximum capacity of GPU memories. The patch size was set to 144 × 144 × 96 voxels in stage I, and 128 × 128 × 128 voxels in stage II. The batch size was set to 2. We use Adam optimizer with initial learning rate of 1e-4. Whenever training loss did not improve by at least $1\times10^{-3}$ within the last 25 epochs, the learning rate was reduced by factor 5. The training was terminated automatically if validation loss did not improve by more than $1\times10^{-3}$ within the last 50 epochs.

To prevent overfitting caused by limited data when training deep neural networks, we applied a series of data augmentation including random rotations, random scaling, elastic deformations, gamma transformation, adding gaussian noise and random mirroring, to enlarge the dataset for training. Those data augmentations were designed to generate extra convincing clinic images based on existing cases. We did not exaggerate the parameters of data augmentations to keep image distortion under control.

## G. Postprocessing

The coarse prediction calculated by stage I 3D U-Net has the same resolution as the resampled input data. Before generating ROI of kidney from stage I prediction, it is necessary to rescale the predicted image back to the original input shape instead of directly using the resampled shape. After that, we cropped the original CTA scans based on the ROI of kidney provided by rescaled predicted image for stage II prediction.

The stage II prediction is not flawless, especially when it comes to vascular segmentation. we cannot fix the awful continuity of vessels in almost all stage IIB predictions after all kinds of trials. Considering the time-cost and operational efficiency, we decided to simply remove discontinuous vessels if their volume is less than 150 mm3. In some cases, we found that a few peripheral organs with abundant vascularity such as spleen, were misinterpreted as kidney and their vessels were extracted mistakenly. To address this problem, we removed all vessels that has no overlap region with the kidney segmentation predicted in stage I. In the end, the stage II predictions were also rescaled to their original shapes.

Those postprocessing is executed in comparisons of loss function and network structures.

# III. Experiments and results

In the section, we will investigate the best-fitting combination of variables relevant to loss function and network architecture through two evaluation metrics: DSC, and sensitivity (SEN). DSC indicated the overlap extent between predicted foreground voxels and ground truth. SEN, also called the true positive (TP) rate, measures the proportion of the foreground voxels that are correctly classified. Specificity (SPE) and accuracy (ACC) are excluded because the background is much larger than foreground, leading their values too high too distinguish. All models are trained from scratch and evaluated using five-fold cross-validation. Since we only focus on the fineness of final tissue extraction, the followed experiments were all in stage II.

## A. Comparisons of loss function

The dice part of our proposed loss function has one variable, which determined the proportion of α and β in formula. We sat a group of paired α and β, from α 0.1 β 0.9, to α 0.9 β 0.1. **Fig. 6** shows how the average DSC and SEN of the proposed network changing on stage II dataset with different recipes.

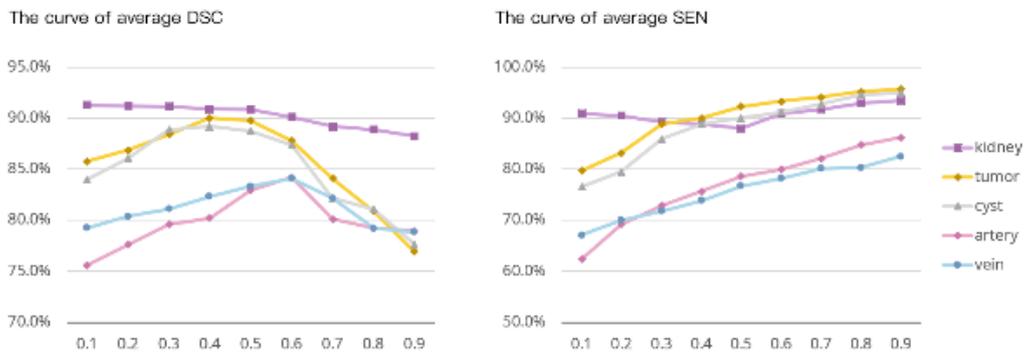

**Fig. 6** The curve of average DSC and SEN. The x-axis represents α value, and β=1-α.

Just like our assumption, there is a general trend of shifting to higher sensitivity when α is growing larger in all classes. Meanwhile, a small rise of DSC is observed at the start of the curve, but soon it dropped when α over 0.5. The peck of DSC curve appears at α 0.4 β 0.6 in stage IIA and α 0.6 β 0.4 in stage IIB. Generally, we should choose the best DSC recipe as the optimal strategy. However, when extracting vessels, the SEN value also needs to be considered.

**Fig. 7** illustrates 2 cases slice image and 3D view with different α and β. it shows how the parameters effect the vessel extraction more visually. As α grows, the sensitivity increases, more vessels were extracted, some are not labeled in ground truth, resulting in the DSC drops when α is too high. But Those extra vessels could be potential region where human cannot recognize. However, higher α is not necessarily better. As α grows, the area of existing vessels also expands gradually, resulting in larger vessel diameter than ground truth. Meanwhile, more tiny discontinuous vessels appear, we cannot confirm whether they are too thin to be continuous or just texture noise being extracted wrongly.

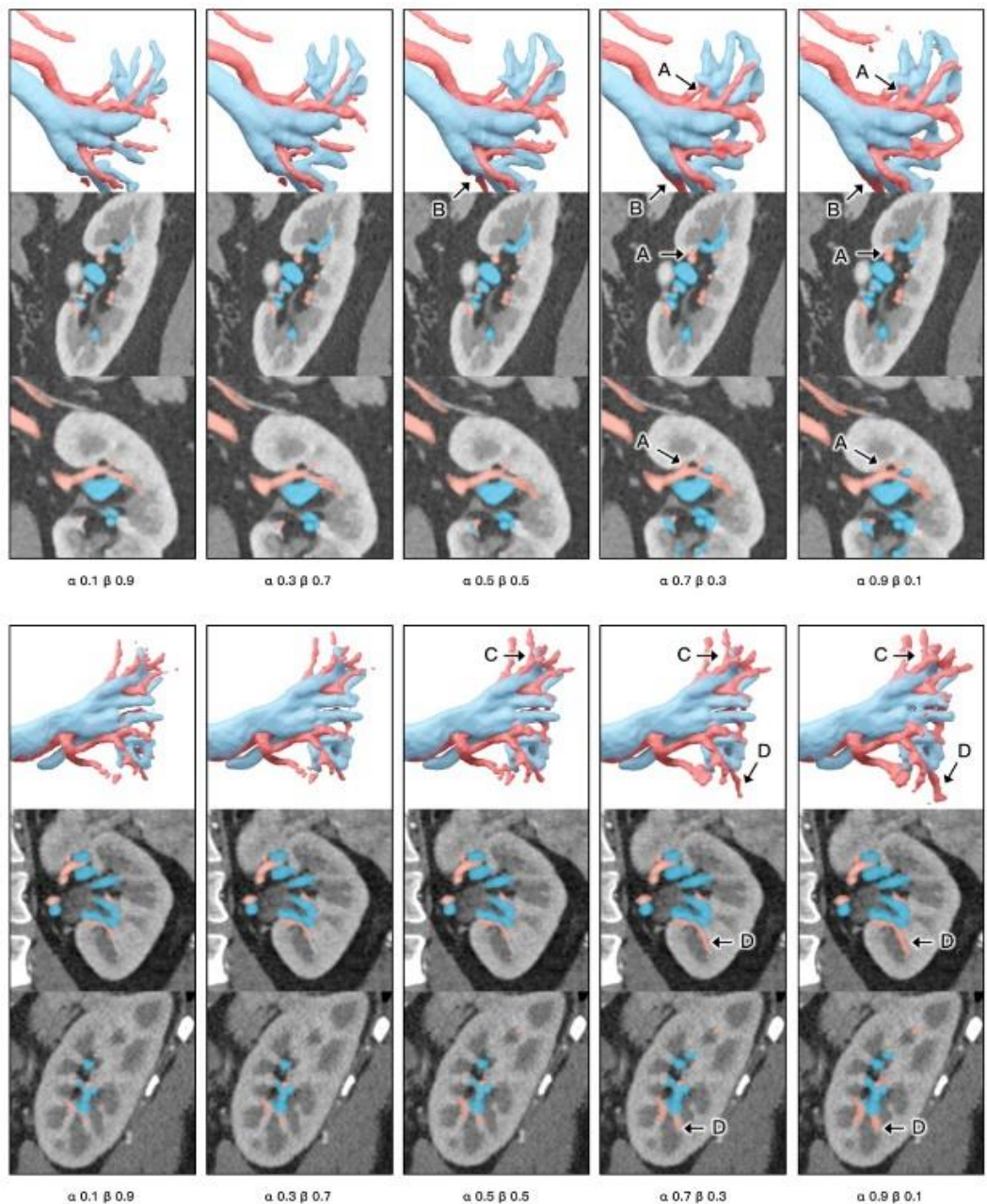

**Fig. 7** the predictions of 2 cases with different α and β. A, B, C, D are the extra vessels, as α grows.

Taking the above observations into consider, we choose α 0.4, β 0.6 for stage IIA, and α 0.7, β 0.3 for stage IIB to maximize the performance of each class. The average DSC value and SEN value of all tissue were 86.5% and 86.0% respectively with this combination of setting.

We also validated the performances of 3 different hybrid strategy of loss function: dice only, dice + cross entropy (CE), and dice + focal. The α and β were set to 0.5 in all dice loss. With average DSC of each class listed in **Table I**, we noticed a general improvement when adopting a hybrid strategy rather than dice loss only. When comprising dice + focal with dice + CE, the former shows a slight advantage in most classes. As we expected, the dice + focal behaves better than dice + CE when it comes to imbalance class situation. Therefore, we choose the dice + focal as our optimal loss function.

**Table I**

The DSC of 3 different hybrid strategy per class.

|  | kidney | tumor | cyst | artery | vein |
| --- | --- | --- | --- | --- | --- |
| dice only | 89.8 | 79.0 | 82.8 | 79.2 | 81.2 |
| dice + CE | 90.7 | 86.1 | 83.3 | 80.1 | 82.6 |
| dice + focal | 90.9 | 89.8 | 88.3 | 82.9 | 83.3 |

Finally, we compared the weighted and non-weighted loss function to prove the importance of modulating the imbalance classes. As it showed in **Fig. 9**, the weight factor dramatically improved the DSC of small objects like tumor, cyst, and vessels by 8-20%.

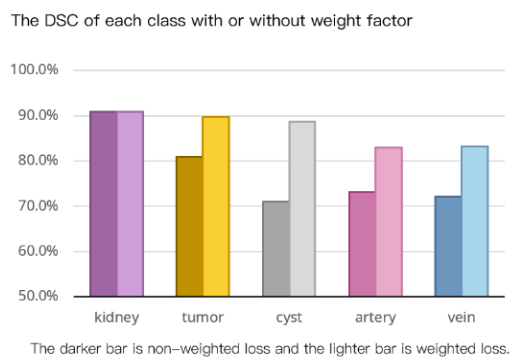

**Fig. 9** The DSC of each class with or without weight factor.

## H. Comparisons of network structures

3D U-Net is characterized by the symmetrical down-sampling path and up-sampling path. The down-sampling path consists of a certain number of encoder blocks and down-sampling blocks in-between. And the up-sampling path consists of the same number of decoder blocks and up-sampling blocks in-between symmetrically. Those 4 kinds of blocks are replaceable, making the 3D U-Net a more flexible architecture. We investigated 4 design of 3D U-Net architectures with same number of feathers and levels of down-sampling, but different in block type:

**Vanilla**, Vanilla 3D U-Net with max-pooling down-sampler, the encoder and decoder blocks are 3D convolutional layers, while the down-sampling blocks are traditional 3D max-pooling layers.

**Conv Down**, 3D U-Net with convolutional down-sampler, the down-sampling blocks are 3D convolutional layers with 2 strides, the rest remains the same as above.

**Residual**, Residual 3D U-Net with max-pooling down-sampler, the encoder and decoder blocks are residual blocks, while the down-sampling blocks are traditional 3D max-pooling layers.

**Full Residual**, Residual 3D U-Net with residual down-sampler, all blocks are replaced with residual blocks with different strides. it was the adopted network architecture design in our final practice as was described specifically in "network architecture" section.

All models were trained with same parameters on both stage IIA dataset and stage IIB dataset for evaluating the performances on each tissue. We listed the average DSC values of each class with the 4 architecture designs in **Table II**, based on which we have some conclusions as following: Firstly, the Full Residual design beats all the other 3 design by 1-9% improvements in average DSC on both datasets. Secondly, max-pooling down-sampler performs worse than residual down-sampler or convolutional down-sampler generally, when comparing Conv Down design to Vanilla design, or Full Residual design to Residual design. Thirdly, the Full Residual design has limited advantage over Residual design on vascular system segmentation.

Based on conclusions above, we choose the Full Residual Design as our optimal network architecture.

**Table II**

The DSC of 4 architecture designs per class.

The DSC of 4 architecture designs per class (%)

|  | kidney | tumor | cyst | artery | vein |
|---|---|---|---|---|---|
| Vanilla | 80.1 | 73.4 | 74.0 | 73.8 | 75.6 |
| Conv Down | 85.5 | 74.7 | 78.1 | 75.8 | 77.2 |
| Residual | 89.4 | 79.1 | 80.7 | 78.8 | 80.9 |
| Full Residual | 89.8 | 79.0 | 82.8 | 79.2 | 81.2 |

## I. *Comparing optimized annotations to original annotations*

In the previous description, we mentioned a non-traditional data process step: optimizing the original vessels annotations in stage IIB dataset.

This optimization is planned to improve the vessels extraction performance based on theory that the quality of annotations has great influence on the result of deep learning method. To prove that we add a new comparison between using optimized annotations and using original annotations when training the models with proposed network. All other training variables stay the same, except datasets.

As it showed in **Fig. 10**, the optimized dataset is superior to the original dataset in all evaluation metrics, whether on arteries or on veins. The former receives 2-5%, 1-2% higher value in

DSC and SEN, respectively. Although, some additional vessels extracted are controversial on authenticity, we got richer and more continuous vessels after adopting the optimization.
There is no doubt that the optimized dataset performs better because we defined the optimized annotations as our ground truth through all evaluations. The comparison, such as it is, has great significance for demonstrating a practical improvement when deal with imperfect manual annotations.

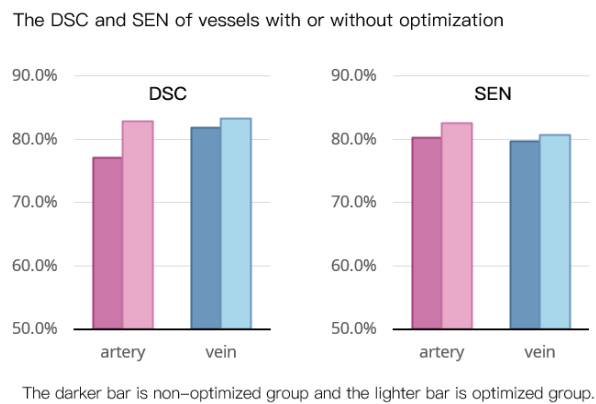

**Fig. 10** The DSC and SEN of vessels with or without optimization.

## J. 3D visualization on renal tumor and its arterial supply

In clinic practice, the PN resection range and position are confirmed based on the spatial relationship between renal tumor and renal artery branches. Through our full-automatic segmentation method, we can obtain all concerned anatomies from abdominal CTA scans for preoperative evaluation. **Fig. 12** presents two 3D visualizations of the renal tumors and their nearest artery branch generated from our models' predication image by software Blender (version 2.83).

Based on those 3D visualizations, we can establish a clear route of tumor's arterial supply in most cases. It can be the guidance of artery clamping that performed in PN surgery. Unfortunately, the blood supply route is not clearly enough to rebuild in few extreme cases. This may be caused by the poor image quality, the improper scan moment for artery phase, or the multi-branches supply. Even so, the general path of artery benches is useful enough for urological surgeon during planning. The fine detail tissue models also can be used in many advanced CAD technologies for further morphological analysis, our visualization is just a simple demonstration.

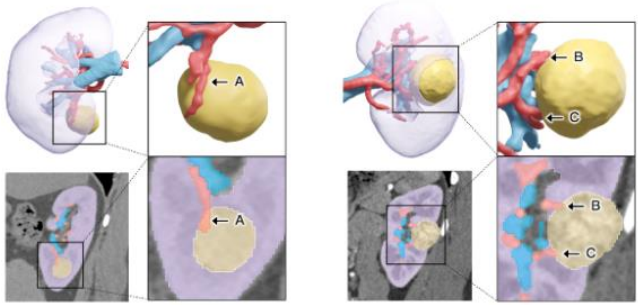

**Fig. 11** 3D visualization of 2 cases. A, B, C are the artery entry point of tumor.

## IV. Discussion

Neural network algorithm is gradually taking center stage in medical segmentation with the rapid development of deep learning technique in the last two decades. They were not only proved preceded over traditional methods in many aspects, but also filled the void on complex tissue extraction which traditional methods cannot handle. In this paper, we provide a robust renal anatomies extraction method only from CTA scan image for PN preoperative assessment. Our proposed automatic algorithm is a representative practice of complex medical segmentation task fully depend on deep learning technique, and proves, once again, the superiority of neural network algorithm through experiments.

With only few annotated data, our model is power enough to guarantee the high accuracy of segmentation. Even when dealing with severe imbalance class such as cysts or vessels, the performance is still not showing a distinct disadvantage. The average DSC of kidney, tumor, cyst, artery, and vein, were 90.9%, 90.0%, 89.2%, 80.1% and 82.2% respectively. This is credited to our loss function strategy from three aspects: First, the proper weights of FP and FN in dice loss increases the sensitivity of tiny objects, leading to good performance on average DSC. Second, the dice loss we adopted is combined with focal loss, which ensuring more effort on poorly classified tissue. Third, an extra weight is added to the loss function to rebalance the occurrence frequency of tissues.

As we have learned, 3D U-Net is a hardware dependent algorithm, due to the large amount of GPU memory usage. To avoid "out of memory" (OOM) issue, steps are taken toward minimizing the memory consumption while still have enough space for building dense network. We addressed this problem by utilizing the "coarse-to-fine" workflow for tissue extraction. In our practice, this strategy is capable for huge shape volumetric data like over 1000 slices with 0.1 mm thickness, while still shows great robustness of generating full resolution prediction labels. And its scope of applications could be extended to most medical tasks with large input data. We approved that the multi-stages workflow may become a convention design for volumetric prediction when the hardware reaches its bottleneck.

Meanwhile, we found that the manual vessels annotations for training are far from perfection. Especially when the vessel is not going along any orthogonal plane (coronal plane, sagittal plane, and horizontal plane) or vessel is too thin to recognize for human eyes. That is because our understanding of spatial structure is mostly rely on two-dimensional images projected onto the retina in daily life. Precise recognition from volumetric image is not our nature. So, we cannot ensure manual segmentations by imaging specialist are free from error, let alone dealing with vessels. Thankfully, we have solution to this problem: the optimization of vessel annotation by semi-trained model under the supervision of experts. In our experiment of comparison between non-optimized and optimized, we draw the conclusion that optimized annotation significantly improves vessels extraction performance in many ways. Inspired by the application in vascular system, we realized that this trick is practical useful to rapidly create labeled medical image for training. It is widely believed that deep learning method on medical is suffered from small dataset, resulting from the difficult of manual segmentation. The number of cases in medical dataset for training is from tens to hundreds, but rarely more than a thousand, particularly in volumetric data such as CT scans. Therefore, we can use the semi-trained model to generate coarse predictions. Those predictions can become the fundamental images for manual segmentation, either regarding as refer or redrawing directly on them. Despite its merits, our proposed method also has many limitations which cannot be ignored. Most limitation is brought by quality of dataset since deep learning's performance depends on it. First, in some extreme cases, the malformed renal may causes unconventional cropping for stage II prediction, resulting in failure overall. It is because limited quantity of stage II dataset does not cover sufficient extreme cases including all types of malformation. This systematic drawback can only be beaten by increasing the dataset capacity which consumes huge amounts of manpower. Comparing to other state-of-art technology, Wang et al. Tensor Cut for instance [11], our algorithm is largely confined to annotation step. Both the quality and the quantity of stage II dataset limit the height of the performance. Even with our proposed "optimizing vessel annotation" method, we still suffer from heavy workload of distinguishing unlabeled vessels form predictions.

Second, CTA images are scanned in arterial phase, from 15-25 seconds after contrast administration[19]. The ideal coverage for renal CTA is between the dome of the diaphragm and the distal portion of the common iliac arteries[20], and renal cortex and veins should not be included. However, the real coverage is uncertain in clinical practices, due to the narrow temporal window. Strictly speaking, many images in stage II from source KiTS19 are not in arterial phase, more like delayed arterial phase or early venous phase. So, the numerical difference between veins and arteries may not great enough to distinguish, adding more difficult to the annotation work for human, leading to less stability of prediction in terminal branches.

## V. Conclusions

In summary, we presented an automatic extraction method for entire renal anatomies from CTA image to assist the preoperative evaluation of PN or any other CAD technique applications of kidney surgery. We chose 3D U-Net as the benchmark and proposed our network for the segmentation task. To avoid OOM issue, we used the coarse-to-fine workflow by applying our deep learning network at two levels of resolution. To handle the imbalance classes and missing annotations, we adopted a weighted hybrid loss function to give equal consideration of every tissue. To complete the defective vessel annotations in original dataset, we optimized the annotations by merging semi-trained model's prediction and original image under supervision. The experiments demonstrated that our algorithm was extremely excellent to extract the renal anatomies in many ways. It was also a significant attempt to segment thin vessels fully based on deep learning technique and proposed several improvements for its performance. This is predictable that with sufficient representative cases for training, the robustness of our workflow will reach to a new high level.

Although, we achieved entire anatomies extraction using automatic program, its practicability is still limited in real clinical situations at present. 3D arteriogram reconstructed images directly rendered from CTA is useful enough to select the right artery for clamping during PN [21]. Our method only shows a small advantage on complicated situation. But if we look further ahead, the digital renal models we generated can be utilized in many other scenarios, not only for PN surgery planning. For example, they can be used in cropping the tumor region for pathological classification based on texture. Moreover, they can be used in labeling the anatomies on the screen of augmented reality equipment during kidney surgery. Even further, they can be used in tissue recognition of future full-automatic surgical robot. We believe that the digitization of renal anatomies will play the fundamental role in advanced urological surgical techniques in the next two decades.

# VI. References

[1]   I. S. Gill *et al.*, "Comparison of 1,800 Laparoscopic and Open Partial Nephrectomies for Single Renal Tumors," *The Journal of Urology*, vol. 178, no. 1, pp. 41–46, Jul. 2007, doi: 10.1016/j.juro.2007.03.038.

[2]   J. Wang *et al.*, "The role of three-dimensional reconstruction in laparoscopic partial nephrectomy for complex renal tumors," *World J Surg Oncol*, vol. 17, Sep. 2019, doi: 10.1186/s12957-019-1701-x.

[3]   L.-M. Su, B. P. Vagvolgyi, R. Agarwal, C. E. Reiley, R. H. Taylor, and G. D. Hager, "Augmented Reality During Robot-assisted Laparoscopic Partial Nephrectomy: Toward Real-Time 3D-CT to Stereoscopic Video Registration," *Urology*, vol. 73, no. 4, pp. 896–900, Apr. 2009, doi: 10.1016/j.urology.2008.11.040.

[4]   Daw-Tung Lin, Chung-Chih Lei, and Siu-Wan Hung, "Computer-aided kidney segmentation on abdominal CT images," *IEEE Transactions on Information Technology in Biomedicine*, vol. 10, no. 1, pp. 59–65, Jan. 2006, doi: 10.1109/TITB.2005.855561.

[5]   G. Yang *et al.*, "Automatic kidney segmentation in CT images based on multi-atlas image registration," in *2014 36th Annual International Conference of the IEEE Engineering in Medicine and Biology Society*, Aug. 2014, pp. 5538–5541, doi: 10.1109/EMBC.2014.6944881.

[6]   H. Song, W. Kang, Q. Zhang, and S. Wang, "Kidney segmentation in CT sequences using SKFCM and improved GrowCut algorithm," *BMC Syst Biol*, vol. 9, no. Suppl 5, p. S5, Sep. 2015, doi: 10.1186/1752-0509-9-S5-S5.

[7]   H. Akbari and B. Fei, "Automatic 3D Segmentation of the Kidney in MR Images Using Wavelet Feature Extraction and Probability Shape Model," *Proc SPIE*, vol. 8314, p. 83143D, Feb. 2013, doi: 10.1117/12.912028.

[8]   K. Sharma *et al.*, "Automatic Segmentation of Kidneys using Deep Learning for Total Kidney Volume Quantification in Autosomal Dominant Polycystic Kidney Disease," *Sci Rep*, vol. 7, May 2017, doi: 10.1038/s41598-017-01779-0.

[9]   E. Gibson *et al.*, "Automatic Multi-organ Segmentation on Abdominal CT with Dense V-networks," *IEEE Trans Med Imaging*, vol. 37, no. 8, pp. 1822–1834, Aug. 2018, doi: 10.1109/TMI.2018.2806309.

[10] C. Bauer, T. Pock, E. Sorantin, H. Bischof, and R. Beichel, "Segmentation of interwoven 3d tubular tree structures utilizing shape priors and graph cuts," *Medical Image Analysis*, vol. 14, no. 2, pp. 172–184, Apr. 2010, doi: 10.1016/j.media.2009.11.003.

[11] C. Wang *et al.*, "Tensor-cut: A Tensor-based Graph-cut Blood Vessel Segmentation Method and Its Application to Renal Artery Segmentation," *Medical Image Analysis*, p. 101623, Dec. 2019, doi: 10.1016/j.media.2019.101623.

[12] Q. Huang, J. Sun, H. Ding, X. Wang, and G. Wang, "Robust liver vessel extraction using 3D U-Net with variant dice loss function," *Computers in Biology and Medicine*, vol. 101, pp. 153–162, Oct. 2018, doi: 10.1016/j.compbiomed.2018.08.018.

[13] W. Huang *et al.*, "Coronary Artery Segmentation by Deep Learning Neural Networks on Computed Tomographic Coronary Angiographic Images," in *2018 40th Annual International Conference of the IEEE Engineering in Medicine and Biology Society (EMBC)*, Jul. 2018, pp. 608–611, doi: 10.1109/EMBC.2018.8512328.

[14] O. Ronneberger, P. Fischer, and T. Brox, "U-Net: Convolutional Networks for Biomedical Image Segmentation," in *Medical Image Computing and Computer-Assisted Intervention – MICCAI 2015*, Cham, 2015, pp. 234–241, doi: 10.1007/978-3-319-24574-4_28.

[15] T.-Y. Lin, P. Goyal, R. Girshick, K. He, and P. Dollár, "Focal Loss for Dense Object Detection," *arXiv:1708.02002 [cs]*, Feb. 2018, Accessed: Jun. 12, 2020. [Online]. Available: http://arxiv.org/abs/1708.02002.

[16] P. S. Liu and J. F. Platt, "CT Angiography of the Renal Circulation," *Radiologic Clinics of North America*, vol. 48, no. 2, pp. 347–365, Mar. 2010, doi: 10.1016/j.rcl.2010.02.005.

[17] A. Aghayev, S. Gupta, B. E. Dabiri, and M. L. Steigner, "Vascular imaging in renal donors," *Cardiovasc. Diagn. Ther.*, vol. 9, no. S1, pp. S116–S130, Aug. 2019, doi: 10.21037/cdt.2018.11.02.

[18] Y. Xu *et al.*, "Three-dimensional renal CT angiography for guiding segmental renal artery clamping during laparoscopic partial nephrectomy," *Clinical Radiology*, vol. 68, no. 11, pp. e609–e616, Nov. 2013, doi: 10.1016/j.crad.2013.06.002.